\documentstyle[12pt]{article}
\begin{document}

\def\Sm{{\cal I}^-}
\def\Sp{{\cal I}^+}
\def\Im{I^-}
\def\Ip{I^+}
\def\Jm{J^-}
\def\Jp{J^+}
\def\M{{\cal M}}
\def\C{{\cal C}}
\def\S{\Sigma}
\def\bS{\bar{\Sigma}}
\def\dS{\partial\Sigma}
\def\tD{\tilde{D}}
\begin{flushright} {\small UMDGR--95--019}\\
                   {\small gr--qc/9410023}\\
\end{flushright}

\bigskip

\begin{center}
{\bf \LARGE Topology of Event Horizons\\
and Topological Censorship}\\[2ex]

\bigskip

Ted Jacobson\footnote{jacobson@umdhep.umd.edu}
and Shankar Venkataramani\footnote{cat@wam.umd.edu}\\
\medskip

{\it Dept. of Physics, Univ.
of Maryland, College Park, MD 20742--4111}\\


\begin{abstract}
{\noindent

We prove that, under certain conditions, the topology of
the event horizon of a four dimensional asymptotically flat
black hole
spacetime must be a 2-sphere. No stationarity assumption
is made. However, in order for the theorem to apply, the
horizon topology must be unchanging for long enough
to admit a certain kind of cross section. We expect this
condition is generically satisfied if the topology is
unchanging for much longer than the light-crossing time
of the black hole.

More precisely,
let $M$ be a four dimensional
asymptotically flat spacetime satisfying the
averaged null energy condition, and suppose that
the domain of outer communication $\C_K$
to the future of a cut $K$ of $\Sm$ is
globally hyperbolic. Suppose further
that a Cauchy surface $\Sigma$ for
$\C_K$ is a topological 3-manifold with compact boundary
$\partial\S$ in $M$, and $\S'$ is a compact submanifold
of $\bS$ with spherical boundary in $\S$ (and possibly
other boundary components in $M/\S$).
Then we prove that the homology group $H_1(\Sigma',Z)$ must be finite.
This implies that
either $\partial\S'$ consists of a disjoint union of
2-spheres, or $\S'$ is nonorientable and $\partial\S'$ contains a
projective plane.
Further, $\partial\S=\partial\Ip[K]\cap\partial\Im[\Sp]$, and
$\partial \Sigma$ will be a
cross section of the horizon as long as no generator of
$\partial\Ip[K]$ becomes a generator of $\partial\Im[\Sp]$.
In this case, if $\S$ is orientable,
the horizon cross section must consist of a disjoint union of
2-spheres.}

\end{abstract}
\end{center}



Hawking proved that
each connected component of the event horizon
of a stationary black hole in four dimensional
general relativity has the
topology of a 2-sphere \cite{Hawking}.
His method of proof was to show that
if the horizon had any other topology, and if the dominant energy
condition is satisfied, one could always find a
(possibly marginally) outer-trapped 2-surface outside the horizon.
This would then imply failure of either
the null convergence condition or cosmic censorship,
since these imply that
such a trapped surface must lie {\it inside} an event horizon.

Hawking's theorem was later generalized by Gannon\cite{Gannon}
who proved, replacing stationarity by some weaker assumptions,
that the horizon must be either spherical or toroidal.
Gannon's assumptions include the dominant energy condition and
the assumption that the horizon is smooth to the future
of some slice, which entails in particular that no new generators enter
to the future of that slice.
Quite recently Browdy and Galloway \cite{Gal} have proved,
also without the stationarity
assumption, that a cross section of a horizon must be spherical.
Their assumptions include a null energy condition and the
assumption that no new generators enter the horizon
on or (therefore) to the future of the
cross section. They also prove that the region of
a partial Cauchy surface outside the black hole and inside a
certain ``well-behaved" sphere is compact and simply connected.

Making use of the topological censorship theorem of Friedman,
Schleich, and Witt \cite{FSW}, we prove in this paper
that a cross section of the event horizon must be topologically
a 2-sphere, provided the horizon topology persists for long enough
in a sense that will be made precise below. In the process we obtain
an explicit restriction on the spatial topology of the domain of
outer communication. A somewhat similar argument for the case of
stationary black holes has been given by Chru\'sciel and Wald \cite{CW}.

The basic idea behind our theorem is that topological censorship
forbids a nonsimply connected event horizon since a causal curve
from $\Sm$ to $\Sp$ can be linked with such a horizon and
not deformable to infinity within the domain of outer
communication. More precisely, the idea is the following.
Let $\M$ be an asymptotically flat spacetime and let $K$
be a cut of $\Sm$. We define the
domain of outer communication to the future of $K$ by
$\C_K:=\Ip[K]\cap\Im[\Sp]$.
Suppose $\C_K$ is globally hyperbolic, and that the averaged
null energy condition holds in $\C_K$.
The topological censorship theorem applied to the spacetime $\C_K$
then implies that all causal
curves from $\Sm$ to $\Sp$ in $\C_K$ must be deformable
to a simply connected neighborhood of infinity.
[Otherwise, according to the argument of \cite{FSW},
$\C_K$ would not
be simply connected, and there would be a surface that is outer trapped
with respect to, and visible from, one of the copies of $\Sp$ in the
covering space of $\C_K$. The existence of such a trapped surface would
be inconsistent with the assumed properties of $\C_K$.]

Having assumed that $\C_K$ is globally hyperbolic, it follows
that it is topologically $\S\times R$, where $\S$ is
any Cauchy surface\cite{Waldbook}. Choosing an arbitrary
nowhere vanishing timelike vector field $t^a$ on $\C_K$,
we can identify all leaves of the foliation with one reference leaf,
by flowing along the integral curves of $t^a$. Thus we can represent
$\C_K$ as the product set $\S\times R$.

It seems reasonable at first to expect that topological
censorship would imply that $\S$ is simply connected.
If $\Sigma$ were {\it not}
simply connected, there would be a noncontractible closed curve
$\lambda$ in $\S$
based at infinity. One could
attempt to ``lift" $\lambda$ to a timelike
curve from $\Sm$ to $\Sp$,
that is not deformable to a neighborhood of infinity,
by breaking $\lambda$ at infinity
and pushing one end off to the future and the other end
off to the past.
[We are assuming here that $\Sigma$ has three or more dimensions
since in two dimensions one can have a noncontractible curve that
is nevertheless deformable to a neighborhood of infinity.]
However, in order for such a strategy to work, it is necesary
that one doesn't ``run out of time", and there seems to be
no guarantee that this cannot happen. For example,
if a new horizon forms,
some integral curve(s) of $t^a$ might necessarily have finite
duration. Or, there might be an inflating region in the spacetime
across which $\lambda$ could not be causally lifted.
Thus we have not succeeded in showing that $\S$ is simply connected.
[If the spacetime is assumed to be stationary, these obstructions
do not occur, and one can prove that $\S$ is simply connected,
as shown in \cite{CW}.]

Since the above strategy for proving that $\S$ must be simply
connected meets with obstacles we were forced to come up with a
different strategy. The one we found involves the first homology
group with integer coefficients $H_1(\S,Z)$, rather than the
fundamental group $\pi_1(\S)$,
and it is applicable only in the case of 3+1 spacetime dimensions.
[$H_1(\S)$ is equivalent to
the abelianization of the fundamental group $\pi_1(\S)$ of
$\Sigma$, i.e., the quotient of $\pi_1(\S)$ by its
commutator subgroup.
(Here and hereafter we suppress explicit reference to the
ring of integers $Z$ in the notation for the homology group.)]

In order to employ the topological results of \cite{Hempel}
we must work with {\it compact} topological manifolds with boundary.
Our theorem will refer to a compact
submanifold $\Sigma'\subset\bS$
of the closure of $\S$ with a single 2-spherical boundary
$S$ in $\Sigma$. In general $\Sigma'$ will also possess boundary
components in $\dS\subset M$, but not in $\Sigma$, which will lie in the
event horizon. In case $\bS$ can be compactified by adding
a point at infinity, $\bS\cup\{\infty\}$ can be employed
instead of a submanifold $\Sigma'$.

The basic idea of our proof
is to show that, for any $\Sigma'$ as above,
$H_1(\S')$ must be {\it finite}, because
if $H_1(\S')$ is infinite,
one can ``sew" a timelike curve from $\Sm$ to $\Sp$
through $\C_K$ in such a way that it is necessarily linked
and not deformable to infinity. This would violate the topological
censorship theorem, so we conclude that $H_1(\S')$ is finite.
Now finiteness of $H_1(\S')$ is a weaker restriction on the topology
than is simply connectedness. Nevertheless,
it turns out that the finiteness
of $H_1(\S')$ is enough of a restriction on $\S'$ to imply,
as long as $\S'$ or $\partial\S'$ is orientable, that $\partial\S'$
must be a disjoint union of 2-spheres\cite{Hempel}.
We will argue below that $\partial\S'/\S$ is always a subset of the
event horizon. If in fact it consists of closed cross sections of
components of the
horizon, it then follows that these components are
2-spheres.

In summary, the theorem to be proved is this:\\
\vskip .25cm
\noindent{\bf Theorem}:\quad Let $M$ be
a four dimensional asymptotically flat spacetime
satisfying the averaged null energy condition, let $K$ be a cut of
$\Sm$, and suppose that the domain of outer communication
$\C_K:=\Ip[K]\cap\Im[\Sp]$ to the future of $K$ is globally hyperbolic.
Suppose further that the closure of a Cauchy surface $\S$ for
$C_K$ is a topological 3-manifold with compact boundary
$\partial\S$,
and that $\Sigma'$ is a compact submanifold of $\bS$
with a single 2-spherical boundary $S$ in $\S$ and possibly
other boundary components in $M/\S$.
Then
\begin{quotation}
(1) $H_1(\Sigma')$ is finite;\\

(2) Either $\partial\S'$ consists of a disjoint union of
2-spheres, or $\S'$ is nonorientable and $\partial\S'$ contains a
projective plane; and \\

(3) $\partial\S=
\partial\Ip[K]\cap\partial\Im[\Sp]$ so,
in particular, $\partial\S$ is a subset of the event horizon.
\end{quotation}

\vskip .25cm

Before presenting the proof of this theorem,
let us offer some fuzzy intuitive remarks about
what it is that will have been proved.
Suppose the topology of a one-component
event horizon of a black hole is unchanging to the future of
some slice. Then it seems that generically there will exist a
cut $K$ of $\Sm$ such that $\partial\Ip[K]$ will intersect
the horizon transversly in a closed surface forming a cross
section. Our result then implies that
this cross section must be a 2-sphere.
More generally, it seems plausible that if
the topology of the horizon is merely unchanging for
``sufficiently long", say, much longer than the light crossing time
corresponding to the asymptotic mass of that component, then there
will exist such a cut $K$. If the horizon has more than one
component, then one expects that similar statements would apply.
Further, if a given component is sufficiently well separated
from the other components so that it is surrounded by a fairly flat
region, then one would expect the other components to be irrelevant.

The proof involves the concept of a
properly embedded, two-sided, nonseparating
surface $D$\cite{Hempel}.
A properly embedded surface $D$ is a compact, connected
2-submanifold of a 3-manifold $\S$ with
$\partial D=D\cap\partial \S$.
It is helpful
to visualize the situation using the simple example of $R^3$
minus a solid torus in which case  a disk in
$\S$ whose boundary lies on the inner cycle of the torus serves
as the surface $D$. This, in fact, is the example we generalized
to come up with the strategy of this proof. A more subtle example
is provided by $R^3$ minus a solid knotted torus, in which
case $D$ can be a Seifert surface \cite{OnKnots}
bounded by a knotted cycle on
the torus. An example where $D$ has no boundary is given by
a wormhole $S^1\times S^2$ minus a point,
in which case
$D$ could be an $S^2$ cross section of the wormhole.

We now present our proof that $H_1(\S')$ must be finite.
Suppose on the contrary that $H_1(\S')$ is infinite.
Then $\S'$ contains a two-sided, properly embedded, nonseparating
surface $D$\cite{Hempel}.
Without loss of generality we can assume that $D$ does not
intersect the spherical boundary $S$, since
this boundary can be eliminated by a one-point compactification.
Thus $D$ is properly embedded, two-sided, and nonseparating
in $\bS$ as well.
Because $D$ is two-sided, one
can define an intersection number of a closed curve $\alpha$ with $D$
as the total number of intersections of $\alpha$ with $D$,
each intersection counted as $\pm 1$ according as $\alpha$
passes from $D^-$ to $D^+$ or vice versa, where $D^{\pm}$
are the two sides of $D$. Because $D$ is a properly embedded
surface,
the intersection number of $\alpha$ with $D$ is a topological
invariant.
If $\alpha$ has nonzero intersection number
with $D$, then $\alpha$ is ``linked" and cannot be contracted to a point
or deformed to infinity. The nonseparating property of $D$ will be
invoked below.

Let $\Sigma$ now denote one particular Cauchy surface in the
foliation of $\C_K$, let
$D$ be a two-sided, nonseparating,
properly embedded surface in $\S$, and let
$\tD$ denote the flow of $D$ along the orbits of $t^a$.
We adopt the name ``C-curves" for timelike curves from
$\Sm$ to $\Sp$ in $\C_K$. Define the subset $F_n\subset \C_K/\tD$
as the collection of all points in $\C_K/\tD$
through which there exists
a C-curve whose future half
has intersection number $n$ with $\tD$.
(Equivalently, the projection of the future half curve
into $\S$ along the orbits of $t^a$ has intersection
number $n$ with $D$.)
Note that since $\C_K$ is the product $\S\times R$, the intersection
number of a C-curve with $\tD$ is well defined and is a topological
invariant of C-curves.

It is easy to see that each
$F_n$ is an open set. Furthermore, since any point of $\C_K$
is in $\Im[\Sp]$, we have
$\C_K/\tD=\cup_n F_n$.
$F_0$ is clearly nonempty, since there are points
near spatial infinity through which there must exist C-curves
that do not meet $\tD$. At least one of the other $F_n$'s must
also be nonempty. To see why, consider any point $d$ in $\tD$.
There must exist a C-curve $\delta$ through $d$ which
intersects $\tD$ transversely at $d$.
If the half of $\delta$ to the future
of $d$ has {\it vanishing} intersection number, then
$p$ just
to the past of $d$ on $\delta$ will be in $F_1$.
If the half of $\delta$ to the future
of $d$ has {\it non}vanishing intersection number $m$, then $p$ just
to the future of $d$ on $\delta$ will be in $F_m$.

Since $F_0$ and at least one of the other $F_n$'s are nonempty,
we have shown
that $\C_K/\tD$ can be written as a union of two nonempty open sets,
$\C_K/\tD=F_0\cup(\cup_{n\ne0} F_n)$. Since $\Sm$ is connected
it follows that $\S$ is connected, and removing the nonseparating
surface $D$ cannot change this, so  $\S/D$ is connected and therefore
so is $\C_K/\tD=(\S/D)\times R$.
Thus the two open sets cannot be disjoint, so there must be
a point $p$ in both $F_0$ and $F_m$ for some nonzero $m$.

Topological censorship implies that all C-curves through $p$ are
unlinked with $\tD$, so that the future and past halves of
all C-curves through $p$ must have opposite intersection numbers.
However, since $p$ is in both $F_0$ and $F_m$, we can
therefore find a C-curve through $p$ whose past half has
zero intersection number and whose future half has intersection
number $m$. But this curve has a net nonzero intersection number
$m$, and is therefore linked, which yields a contradiction.
We thus conclude that no surface such as $D$ can exist, and therefore that
$H_1(\S')$ must be finite for any compact submanifold
$\S'\subset\bS$ with 2-sphere boundary in $\S$.

Finiteness of $H_1(\Sigma')$ implies that
either $\partial\S'$ consists of a disjoint union of
2-spheres, or $\S'$ is nonorientable and $\partial\S'$ contains a
projective plane\cite{Hempel}. This establishes the second
claim of the theorem.

Finally, how is $\partial\S$ related to the horizon?
We now argue that $\partial\S=\partial\Ip[K]\cap\partial\Im[\Sp]$,
the third claim of the theorem.
Any point $x$ in $\partial\S$
is also in $\partial\C_K=\partial(\Ip[K]\cap\Im[\Sp])$ and,
since $\S$ is a Cauchy surface for $\C_K$,
$x$ cannot be in the interior of either $\Ip[K]$ or
$\Im[\Sp]$. Therefore, $x$ must lie in
$\partial\Ip[K]\cap\partial\Im[\Sp]$.
Conversely, if $x$ is in
$\partial\Ip[K]\cap\partial\Im[\Sp]\subset\partial\C_K$,
then $x$ must be in the closure of $\S$. Otherwise, there would
exist an open set around $x$ which does not intersect $\Sigma$,
and one could therefore find a timlike curve through a point in
$\C_K$ that does not intersect the Cauchy surface $\S$. Furthermore,
$x$ could not be an interior point of $\S$, since $x\in\partial\C_K$.
Thus $x\in\partial\S$. This completes the proof of our theorem.

The above argument shows that $\partial\S$ is the subset
$\partial\Ip[K]\cap\partial\Im[\Sp]$
of the event horizon. However, we do not know that it
is a union of cross sections of components of the horizon.
With the additional hypothesis that
no generator of $\Ip[K]$ becomes a horizon generator
we can prove that $\partial\Ip[K]\cap\partial\Im[\Sp]$
intersects no horizon generator more than once.
If some generator is intersected in two points $p$ and $q$,
then the later
of the two points $q$ will be joined to $K$ by a segment
of the horizon generator back to $p$ followed by a generator of
$\partial\Ip[K]$ back to $K$. If
no generator of $\Ip[K]$ becomes a horizon generator
then these two curves must form a broken future pointing
null geodesic, which would contradict the fact that
$q$ is in the boundary $\partial\Ip[K]$.
\vskip .5cm

The remainder of this paper consists of several remarks.
\vskip .25cm

$i$)
We have not proved anything about $H_1(\S)$ nor $\pi_1(\S)$
in more than three spatial dimensions. But, even if we could
establish some such result, it would say absolutely nothing
about the topology of the horizon {\it by itself}.
The reason is that, in more than three dimensions,
 {\it any} orientable  manifold
that is the boundary of some other
manifold is also the boundary of a simply connected manifold.
This is because one can always kill the homotopy classes by surgery
without changing the boundary of the manifold \cite{Milnor}.
[To kill the homotopy class of a curve $\lambda$ in an $n$-manifold,
``thicken" $\lambda$ to obtain a solid tube $S^1\times D^{n-1}$,
cut out the solid tube, and replace the solid tube by the simply
connected manifold $D^2\times S^{n-2}$ (which has the same boundary,
$S^1\times S^{n-2}$, as the solid tube.)]
For example $S^1\times D^3$ is nonsimply connected with
nonsimply connected boundary $S^1\times S^2$.
Surgery yields the simply connected manifold $D^2\times S^2$
with the same boundary.

$ii)$ A 3+1 dimensional static black hole with cylindrical
event horizon has recently been found in the presence of a
negative cosmological constant \cite{cyl3}.
This spacetime does not contradict our result because
it is not asymptotically flat in the usual sense, and
there is no simply connected neighborhood of
infinity.
However, although the proof of topological censorship in \cite{FSW}
employed the existence of a simply connected neighborhood of infinity,
it appears likely that the theorem can be extended to spacetimes with
no such neighborhood.
[The spacetime of \cite{cyl3} is asymptotically
anti-de Sitter in the ``transverse" directions rather than
asymptotically flat, so it is not immediately
clear whether or not the topological censorship theorem would
apply to it anyway. Nevertheless, we believe it is not hard to
extend the theorem to cover this asymptotic structure.]
If this is the case,
topological censorship would
still imply that all causal curves from $\Sm$ to $\Sp$ are deformable
to a neighborhood of infinity.
Nevertheless, the presence of a cylindrical horizon
does not prevent this from being true. Any curve that wraps around the
horizon can just be pulled out to wrap around infinity.
Similar comments apply to the recently discussed
2+1 dimensional black holes with
circular event horizons \cite{cyl2}.

$iii$) A 3+1 dimensional static black hole with toroidal event
horizon can be obtained by taking the product of a static
1+1 dimensional black hole with a flat $T^2$.
This spacetime does not contradict either Hawking's theorem
or ours,
for several independent reasons. The 1+1 spacetime has positive
curvature and the transverse space is flat, so one finds that the
effective 4d stress tensor has only negative transverse pressure
components. Thus all energy conditions are violated \cite{Horo}.
Furthermore,
this spacetime is not asymptotically flat in the usual sense.
The expansion of a null congruence orthogonal to a spacelike
cut of null infinity can vanish in this spacetime, so one can not
even argue that a marginally trapped surface must lie inside an event
horizon. Moreover, there is no simply connected neighborhood of
infinity. The last two problems can be circumvented by splicing
on to a standard asymptotically flat metric outside some
radius \cite{GH}. Evidently, however, this cannot be done without
violating the averaged null energy condition, since otherwise
one would have a marginally trapped surface visible from $\Sp$.

$iv$) The recent work of Hughes et. al. \cite{toroid}
demonstrates numerically that a spinning toroid of collisionless
matter can collapse to form a black hole whose event horizon
is initially toroidal and later becomes spherical.
The existence of the temporarily toroidal horizon does not
obviously contradict our theorem since the toroidal topology is not
unchanging to the future of some slice.
However, our theorem implies that whenever a temporarily
nonspherical horizon occurs, it must
not remain nonspherical long enough to permit the existence of a cut
$K$ of $\Sm$ such that the boundary $\partial\S$ of the
Cauchy surface for $\C_K$ is a non-spherical cross section.
(This fact seems closely related to the puzzle raised in remark
($vi$) below.) Evidently,
for any cut $K$, either $\partial\S$ is a spherical cross section
to the future of the torus, or it is spherical but fails to be a cross
section, or $\S\cup\partial\S$ fails to be a 3-manifold with
compact boundary $\partial\S$.

$v$) The topological censorship theorem proved in \cite{FSW}
showed that every causal curve
from $\Sm$ to $\Sp$ must be deformable to lie in a simply connected
neighborhood of infinity. We conjecture that the theorem may be
strengthened to show, with the same assumptions, that
each such causal curve must be deformable to
infinity through a family of causal curves that all
emerge from the same point of $\Sm$. We will call such
deformations ``causal". To establish this conjecture, it would
seem natural to adapt the alternate proof of the topological censorship
theorem, mentioned in \cite{FSW}, which uses
the concept of ``fastest curve" from a given generator of
$\Sm$ to a given generator of $\Sp$.

$vi$) Our conjecture in the preceding remark poses the following
puzzle regarding the temporarily toroidal horizon of Ref. \cite{toroid}.
Consider the one parameter family of null rays that travel along
the axis of symmetry, in particular
the rays that arrive at the location of the spinning toroid
at intermediate times, after the toroidal horizon has formed but
before it has become spherical. It seems at first that
some of these rays will make it back out to infinity but
will {\it not} be causally deformable to infinity, because the
required deformation would necessarily intersect the horizon.
Thus either our conjectured strengthening of the topological
censorship theorem is wrong,
or in fact all these null rays are causally deformable to infinity.

Since the toroidal horizon persists only for a time of order
$M$, a time advance of order $M$ would suffice to disentangle
the ray from the horizon. Perhaps by increasing the impact parameter,
i.e., moving off axis but still coming in from the same point of
$\Sm$, a causal curve could arrive at the spinning toroid before the
formation of the event horizon, and could thus be further causally
deformed to infinity.

\vskip 1cm

We are grateful to D.~Brill, S.~Corley, W.~Goldman, C.~W.~Misner,
J.~Schafer, and J.~Z.~Simon for helpful discussions, and especially
grateful to G.~Galloway, P.~Chru\'sciel, and R.M.~Wald for
comments and corrections on a draft of this paper.
T.~J.~would like to thank the Institute for Theoretical Physics
at the University of Bern for hospitality and
Tomalla Foundation Zurich for support while
this work was being completed.
This work was supported in part by NSF Grant PHY91-12240.


\begin{thebibliography}{999999}
\bibitem
{Hawking}Hawking, S.~W., {\sl Comm. Math. Phys} {\bf 25}, 152 (1972);
Hawking, S.~W.~ and Ellis, G.~F.~R., {\sl The Large Scale Sructure
of Spacetime}, Cambridge University Press (Cambridge, 1973).
\bibitem
{Gannon}Gannon, D., {\sl Gen. Rel. Grav.} {\bf 7}, 219 (1976).
\bibitem
{Gal}Browdy, S. and Galloway, G.J., preprint;
Galloway, G., ``Least area tori, black holes and topological
censorship", to appear in {\sl Contemp. Math.} {\bf 170},
eds. J. Beem and K.L. Duggal (Amer. Math. Soc., Providence, 1994)
\bibitem
{FSW}Friedman, J.~L., Schleich, K. and Witt, D.~M.,
{\sl Phys. Rev. Lett.} {\bf 71}, 1486 (1993).
\bibitem
{CW}Chru\'sciel, P.T. and Wald, R.M., ``On the topology of stationary
black holes", preprint (1994).
\bibitem
{Waldbook}Wald, R.~M., {\sl General Relativity}, The University of
Chicago Press (Chicago, 1984).
\bibitem
{Hempel}Hempel, J., {\sl 3-Manifolds}, Princeton University Press
(Princeton, 1976).
\bibitem
{OnKnots}Kauffman, L.~H., {\sl On Knots}, Princeton University Press
(Princeton, 1987).
\bibitem
{Milnor}Milnor, J., in {\sl Proceedings of Symposia in Pure Mathematics,
volume III: Differential Geometry}, American Mathematical Society
(Providence, 1961).
\bibitem
{cyl3}Lemos, J.~P.~S., gr-qc/9404041.
\bibitem
{cyl2}Ba\~nados, M., Teitelboim, C., and Zanelli, J.,
{\sl Phys. Rev. Lett} {\bf 69}, 1849 (1992).
\bibitem
{Horo}Horowitz, G.~T., in {\sl String theory and quantum gravity
`92}, ed. J.~Harvey et.~al., World Scientific (Singapore, 1993).
\bibitem
{GH}Geroch, R., and Hartle, J.,
{\sl J. Math. Phys.} {\bf 23}, 680 (1982).
\bibitem
{toroid}Hughes, S.~A., et. al.,
{\sl Phys. Rev. D} {\bf 49}, 4004 (1994).
\end{thebibliography}
\end{document}